\documentclass[aps,prb,twocolumn,superscriptaddress]{revtex4-1}
 \usepackage[version=3]{mhchem} 
 \usepackage[T1]{fontenc}       
 \usepackage{graphicx} 
 \usepackage{epstopdf}
 \usepackage{color}
\PassOptionsToPackage{english}{babel}
\usepackage{amssymb,amsfonts,amsmath}
\usepackage{bbold}
\usepackage[normalem]{ulem}
\usepackage{morefloats}
\usepackage{soul}
\usepackage[english]{babel}
\usepackage{multirow}
\usepackage{placeins}

\newcommand{\revision}{\textcolor{black} }

\newcommand{\dir}{.}

\newcommand{\ud}{{\rm d}}

\newcommand{\Rg}{R_g}

\newcommand{\qq}{{\mathbf{q}}}
\newcommand{\hq}{{\mathbf{\hat{q}}}}
\newcommand{\rr}{{\mathbf{r}}}
\newcommand{\vv}{{\mathbf{v}}}
\newcommand{\ff}{{\mathbf{f}}}
\newcommand{\jj}{{\mathbf{j}}}
\newcommand{\II}{{\mathbf{I}}}

\newcommand{\RR}{{\mathbf{R}}}
\newcommand{\VV}{{\mathbf{V}}}
\newcommand{\hmu}{{\hat{\mu}}}
\newcommand{\hL}{{\hat{\Lambda}}}

\newcommand{\matL}{\underline{\underline{\Lambda}}}
\newcommand{\hmatL}{\hat{\underline{\underline{\Lambda}}}}
\newcommand{\matG}{\underline{\underline{g}}}
\newcommand{\matT}{\underline{\underline{T}}}
\newcommand{\vecR}{\underline{\rho}}
\newcommand{\vecM}{\underline{\mu}}

\setstcolor{black}

\begin{document}

\author{Sriteja Mantha}
\affiliation{Institut f{\"u}r Physik, Johannes Gutenberg Universit{\"a}t Mainz, Staudingerweg 9, 55128 Mainz, Germany}
\author{Shuanhu Qi}
\affiliation{Key Laboratory of Bio-inspired Smart Interfacial Science and Technology of Ministry of Education, School of Chemistry, Beihang University, Beijing 100191, China}
\author{Friederike Schmid}
\email{schmidfr@uni-mainz.de}
\affiliation{Institut f{\"u}r Physik, Johannes Gutenberg Universit{\"a}t Mainz, Staudingerweg 9, 55128 Mainz, Germany}

\title{Bottom-up construction of dynamic density functional theories 
for inhomogeneous polymer systems from microscopic simulations} 




\begin{abstract}

We propose and compare different strategies to construct dynamic density
functional theories (DDFTs) for inhomogeneous polymer systems close to
equilibrium from microscopic simulation trajectories. We focus on the
systematic construction of the mobility coefficient, $\Lambda(\rr,\rr')$, which
relates the thermodynamic driving force on monomers at position $\rr'$ to the
motion of monomers at position $\rr$. A first approach based on the Green-Kubo
formalism  turns out to be impractical because of a severe plateau
problem. Instead, we propose to extract the mobility coefficient from an
effective characteristic relaxation time of the single chain dynamic
structure factor.  To test our approach, we study the kinetics of ordering and
disordering in diblock copolymer melts.  The DDFT results are in very good
agreement with the data from corresponding fine-grained simulations. 

\end{abstract}

\maketitle
\section{Introduction}

Inhomogeneous polymer systems assemble into ordered morphologies due to
incompatible interactions between different constituents in the
systems\cite{Hong1981,Schacher2012}. These morphologies have found
applications as thermoplastic elastomers\cite{Lynd2010}, materials for drug
delivery and release\cite{Liechty2010}, gas capture\cite{Galizia2017}, water
purification\cite{Fane2015}, energy conversion\cite{Peng2017a,Peng2017b}, and
also in soft lithography\cite{Black2007}. Understanding the relation between
the molecular features of polymers and the ordered morphologies formed by them
has been a subject of active investigation for a long
time\cite{Bates2017,Polymeropoulos2017,Knychala2017,Chen2017}. An equally
interesting topic is the effect of polymer dynamics on the process of
self-assembly\cite{Fredrickson1996}, e.g., on the kinetics of defect formation
depending on the way a nanostructured polymer material is processed
\cite{Tsarkova2006,Li2014,LiMu2015,Hur2015,Abate2019}. This has lead to
experimental and theoretical investigations to understand the polymer dynamics
in inhomogeneous systems and its effect on the formation of ordered
morphologies. 

Different scattering and reflectometry techniques have been employed to study
the kinetic pathways leading to order-order and order-disorder transitions in
block copolymer
systems\cite{Krishnamoorti2000,Bates1994,Sakurai1996,Jeong2003,Sota2003,Wang2002,Hajduk1994,Schulz1994}.
The same techniques are used to investigate the adsorption dynamics and the
formation of interfaces in an incompatible homo-polymer
blend\cite{Fialkowski2002,Coppee2011,Xu2011,Klein1990,Steiner1990,Guckenbiehl1994,Scheffold1996,Composto2002,Lo2005,Schaefer2017}.
However, the dynamics in inhomogeneous polymer systems involves relaxation
processes occurring over multiple length and time scales. For example, the
molecular features of polymers determine the local rearrangements of chains. On
the other hand, the mesoscopic ordering of polymer chains takes place on length
and time scales which are multiple orders of magnitude higher than the
molecular length and time scales. As a result, finding an experimental
technique that can capture the dynamics over the entire spectrum of length and
time scales is an extremely involved task. Dynamic density functional theory
(DDFT)
\cite{Fraaije1993,Kawasaki1987,Kawasaki1988,Fraaije1997,Muller2005,Kawakatsu1999}
or the dynamic self-consistent field theory have been promoted as a theoretical
alternative to study the polymer dynamics on the relevant mesoscopic length and
time scales. 

In a DDFT, the dynamics of an inhomogeneous polymer system is described
by a diffusive equation in the monomer densities
\begin{equation}
\frac{\partial \rho_\alpha\left(\rr,t\right)}{\partial t} 
  = \sum\limits_{\beta} \nabla_r \left[\int \ud \rr' 
    \Lambda_{\alpha \beta} \left(\rr,\rr'\right)\nabla_{r'}
     \mu_\beta \left(\rr',t\right)\right]
\label{eq:Intro}
\end{equation}

Here, $\rho_{\alpha}\left(\rr,t\right)$ is the density of monomers of type
$\alpha$, $\Lambda_{\alpha \beta} \left(\rr,\rr'\right)$ is the mobility matrix
and $(- \nabla_{r'} \mu_{\beta}\left(\rr',t\right))$ a local thermodynamic force
acting on monomers of type $\beta$.  The matrix $\matL\left(\rr,\rr'\right)$
relates the monomer density current to the thermodynamic driving force
\cite{Binder1987} and depends on the monomer-monomer correlations in the
system.  The field $\mu_{\beta}(\rr,t)$ can be interpreted as a local chemical
potential for {\em unconnected} monomers of type $\beta$ and is derived from a
free energy functional $F$, (i.e., $\mu_\beta\left(\rr',t\right) =\delta
F/\delta \rho_\beta(\rr',t)$), which is typically taken from self-consistent
field (SCF) theory.  Since $\rho_{\alpha} \left(\rr,t\right)$ are
coarse-grained quantities, their dynamic evolution equations describe the
kinetics in the system on mesoscopic scales.  A typical SCF
theory\cite{Schmid1998,Matsen2002, Fredrickson} for polymers retains
microscopic information on the chain architectures.  This combination of
mesoscopic and microscopic aspects makes DDFT a promising technique in the
pursuit of studying polymer dynamics in an inhomogeneous system. DDFT has been
used to explore the kinetic pathways for micelle to vesicle transition in 
micellar solutions\cite{He2006,He2008}, morphological transitions in 
diblock copolymer melts \cite{Wang2011,Morita2002,Morita2001} and also scaling
laws for the polymer inter-diffusion during interfacial broadening in 
polymer blends\cite{Reister2001,Yeung1999,Qi2010,Reister2003}. DDFT models have
also been extended to study the effects of
hydrodynamics\cite{Maurits1998,Honda2008} and
reptation\cite{Maurits1997,Shima2003}.  Recent investigations have also used
DDFT in conjunction with the string method \cite{E2002} to determine the mean
free-energy path for pore formation and rupture in cell
membranes\cite{Ting2011}.  

Although DDFT has significantly advanced our understanding of polymer dynamics,
it suffers from the problem that DDFT models are typically constructed in an ad
hoc manner.  The dynamics of polymers is well-known to be governed by
relaxation processes on multiple time scales\cite{Doi}. When projecting the
dynamical equations for monomer coordinates onto a dynamical equation for
densities such as Eq.\ (\ref{eq:Intro}) in a systematic manner, e.g., using the
Mori-Zwanzig formalism\cite{Zwanzig1961, Mori1965}, this invariably results in
a generalized Langevin equation with a memory kernel \cite{Wang2019}. In DDFT,
the memory kernel is replaced by one single, time independent (but nonlocal)
effective mobility function. This greatly increases the computational efficacy
of the resulting coarse-grained model, however, the optimal way to choose such
an effective mobility is not clear.

Currently, all approaches in the literature are based on heuristic assumptions.
For chains in the Rouse regime, these approximate schemes can broadly be
categorized into local and nonlocal approaches\cite{Qi2017}. In the local
approach, monomers are assumed to diffuse in the system independent of each
other. In the nonlocal approaches, polymers are assumed to diffuse as a whole.
These approximations significantly reduce the complexity in handling the DDFT
equation. However they come with their own caveats. Most importantly, it was
found that the choice of DDFT approach may influence the pathways of
self-assembly that are observed in DDFT calculations. One example is the
dynamics of vesicle formation from homogeneous nucleation, where nonlocal DDFT
calculations predicted the existence of competing pathways of self-assembly
\cite{He2006, He2008} (which was then confirmed both by
experiments\cite{Han2010,Wu2018,Ianiro2019} and simulations\cite{Zeng2016,
Meiling2014,Xiao2012,Huang2009}), whereas only one pathway was present in local
DDFT simulations\cite{Sevink2005}. Moreover, local DDFT calculations greatly
overestimate the frequency of vesicle fusion events \cite{Zhang2011}, which are
largely suppressed in nonlocal DDFT simulations \cite{He2006a,Heuser2017}
consistent with experiments \cite{Wu2018}. When comparing to particle-based
simulations, local DDFT calculations generally tend to overestimate the speed
of structure formation, and nonlocal DDFT calculations tend to underestimate it
\cite{Reister2003, Zhang2011,Qi2017}.

It should be noted that none of these approaches incorporate knowledge on the
microscopic dynamics in the underlying polymer dynamics. In recent years,
bottom-up coarse-graining techniques have become increasingly popular in
materials science, where coarse-grained models are constructed from
fine-grained simulations in a systematic manner.  Examples are techniques for
deriving effective potentials in coarse-grained
models\cite{Lyubartsev1995,Reith2003,Florian2002, Karimi2012,Gooneie2017} or
effective friction coefficients\cite{Deichmann2018} or even memory
kernels\cite{Schnurr1997, Shin2010, Carof2014, Li2015, Li2017,
Jung2017,Jung2018,Meyer2020} in dynamical equations.  Since SCF models bridge
between microscopic and the mesoscopic length scales, it should be possible to
apply similar ideas for the construction of DDFT equations in order to improve
their predictive capabilities.

In this article, we explore two physically motivated bottom-up construction
schemes for determining DDFT mobility functions $\Lambda\left(\rr,\rr'\right)$
from microscopic simulations. In the first approach, we follow a classical
approach to this type of problem and consider the Green-Kubo relation
\cite{Kubo,Hansen,Kubo1996} that relates $\Lambda\left(\rr,\rr'\right)$ to an
integral over an appropriate current-current time correlation function.
Unfortunately, the result turns out to be not very useful, for reasons that we
shall discuss below. In a second approach, we therefore propose to extract
$\Lambda\left(\rr,\rr'\right)$ from the characteristic relaxation time of the
dynamic structure factor of single chains. 

To test our approach, we study two related problems: The first is the dynamics
associated with the formation of the lamellar structure in diblock copolymer
melts, the second is the relaxation of a lamellar structure into a homogeneous
state. We specifically choose these problems because existing local and
non-local DDFT schemes are known to significantly under- or overestimate the
time scales of (dis)ordering in comparison to fine grained simulations of the
same systems.  We show that the bottom-up constructed DDFT models are able to
capture both the global dynamics and the relaxation due to local
rearrangements of the chain at the relevant length scales. This significantly
improves the DDFT predictions for the above listed problems. 

The rest of the manuscript is organized as follows: In the next section, we
first introduce the general framework of DDFT theory and briefly describe the
Ans\"atze for mobility functions that have been proposed in the literature.
Then we present and discuss our two bottom-up approaches. Finally, in the
fourth section, we apply the approach to the study of ordering and disordering
in diblock copolymer melts. We conclude with a summary and an outlook.

\section{General framework of DDFT}

The dynamic density functional theory is an extension of the classical density
functional theory, where the equilibrium free energy of a many-body system is
expressed as a functional of coarse-grained field variables, the density
fields\cite{Hansen,Evans1979,Mermin1965}. A mathematical basis for this
formalism is provided by the Hohenberg-Kohn theorem
\cite{Hohenberg1964,Kohn1965}. Here we consider polymer systems with
different types of monomers $\alpha$, hence our free energy functional depends
on several fields, $F(\{\rho_{\alpha} \})$.  In practice, we will use the
functional provided by the self-consistent field (SCF) theory
\cite{Schmid1998,Matsen2002,Fredrickson}, which is a mean-field approach. 

The objective of the DDFT is to construct a physically motivated scheme for the
dynamical evolution of the microscopic densities, based on the given static
functional. Such a scheme is expected to drive the system along a path of low
free energy, with meaningful dynamic information, in order to reach the
equilibrium state or at least a metastable minimum of $F$. Since the density is
a conserved field, its longest-wavelength Fourier components are slowly
relaxing variables \cite{Fredrickson}. This motivates the construction of a
diffusive equation that involves the dynamic evolution of density fields only,
resulting in so-called model B dynamics according to the classification of
Hohenberg and Halperin \cite{Hohenberg1977}.

A simple popular Ansatz is to assume the linear instantaneous form 
\begin{equation}
\frac{\partial \rho_\alpha\left(\rr,t\right)}{\partial t} 
  = \nabla_r \sum_\beta\int \ud\rr' \Lambda_{\alpha\beta}
    \left(\rr,\rr'\right)\nabla_{\rr'}\mu_\beta\left(\rr',t\right)
\label{eq:DDFT0}
\end{equation}
with $\mu_{\beta} (\rr,t) = \delta F/\delta \rho_{\beta}(\rr, t)$. The mobility
function $\Lambda_{\alpha \beta}(\rr,\rr')$ relates the density current of the
monomer $\alpha$ at position $\rr$ to the thermodynamic driving force ($-
\nabla \mu_\beta$) on the monomer $\beta$ at position $\rr'$. In the present
paper, we will consider single-component homopolymer or copolymer melts with
average monomer density $\rho_0$, and assume that all chains have equal length
$N$.  Furthermore, to simplify the notation, we will often use reduced
quantities $\phi_{\alpha}= \rho_\alpha/\rho_0$, $\hmu_\beta = \frac{N}{\rho_0}
\delta F/\delta \phi_\beta = N \mu_{\beta}$, and $\hL = \Lambda/\rho_0 N$,
which allows us to rewrite (\ref{eq:DDFT0}) as
\begin{equation}
\frac{\partial \phi_\alpha\left(\rr,t\right)}{\partial t} 
  = \nabla_r \sum_\beta\int \ud \rr' \hL_{\alpha\beta}
    \left(\rr,\rr'\right)\nabla_{r'}\hmu_\beta\left(\rr',t\right).
\label{eq:DDFT}
\end{equation}

We note that the instantaneous assumption is questionable in polymeric systems,
which are known to exhibit memory effects \cite{Doi}, as already discussed in
the introduction. In DDFT, one implicitly assumes that the memory kernel can
be replaced by a simple, time-independent (but not necessarily local) function.
A second important approximation, which is typically made in polymeric DDFT
approaches and which we will also adopt here, is a mean-field approximation: In
the spirit of the SCF theory which provides the static density functional $F$,
polymers are assumed to move independently in an external field provided by the
other polymers. This field may include hydrodynamic flows and even
entanglements, but only in an averaged sense. Hence the mobility function
$\Lambda$ describes the mobility of individual chains. It includes effects of
intrachain monomer correlations, but not those of interchain
correlations. From Eq.\ (\ref{eq:DDFT0}), one can thus extract a mobility
function {\em per chain}, given by $\Lambda^{(s)} = \Lambda N/\rho_0 = \hL
\:N^2$.

For melts in the Rouse regime (i.e., chains are non-entangled), three types of
Ansatz for the mobility coefficients have been proposed in the literature:

{\it{(i) Local coupling scheme:}} 
In this approximation, monomer beads are assumed to diffuse
independently of each other with the mobility $D_0/k_B T$\cite{Fraaije1993}. 
This leads to the following expression for
$\hL_{\alpha\beta}\left(\rr,\rr'\right)$:

\begin{equation}
\hL_{\alpha\beta}^{\mbox{\tiny Local}}\left(\rr,\rr'\right)
= \frac{D_0}{N k_B T} \phi_\alpha\left(\rr\right)
   \delta_{\alpha\beta} \delta\left(\rr-\rr'\right) 
\label{eq:Local}
\end{equation}

{\it{(ii) Chain coupling schemes:}} These approaches assume that the
internal structure of the polymer chain relaxes on a time scale much faster
than the collective motion of the chain. As a consequence, the polymer chains
are assumed to diffuse as a whole with the mobility $D_c=D_0/N$. For this case,
Maurits et al have derived the expression \cite{Maurits1997}
\begin{equation}
\hL_{\alpha \beta}^{\mbox{\tiny Chain}}(\rr,\rr',t)
= \frac{D_c}{k_B T} P_{\alpha \beta}(\rr, \rr', t) / \rho_0 N
\label{eq:FullChain}
\end{equation}
where $P_{\alpha \beta}(\rr, \rr',t)/\rho_0 N$ is the pair correlation of 
monomers $\alpha$, $\beta$ on the same chain at position $\rr$ and $\rr'$,
normalized to the integral one. Within the SCF theory, this quantity
can be calculated exactly using a scheme proposed earlier by two of
us \cite{Qi2017}. Further approximations have been proposed, such as
the external potential dynamics (EPD) approximation (not discussed here)
and the Debye approximation, which approximates $P_{\alpha \beta}/\rho_0$
by the pair correlations of ideal Gaussian chains, i.e., the Debye 
correlation function\cite{Doi}
\begin{equation}
\hL_{\alpha\beta}^{\mbox{\tiny Debye}}\left(\rr,\rr'\right) =
 \frac{D_c}{N k_B T} g_{\alpha\beta}\left(\rr-\rr'\right) 
\label{eq:Debye}
\end{equation}
Analytical expressions are available for the Fourier representation of $g(\rr -
\rr')$. For example, for diblock copolymers, one obtains
\cite{Rubenstein,Fredrickson} 
\begin{eqnarray}
g_{\alpha\alpha}\left(q\right)& =& N f_D\left(h_\alpha,x\right)
\\
\nonumber 
g_{AB} &=& \frac{N}{2} \left\{ f_D\left(1,x\right) 
   -f_D\left(h_A,x\right) -f_D\left(h_B,x\right)\right\}
\label{eq:gD}
\end{eqnarray}
where $x=q^2\Rg^2$, $h_\alpha$ is the fraction of block $\alpha$, and 
\mbox{$f_D\left(h,x\right):=\frac{2}{x^2}\left(hx+e^{-hx}-1\right)$}
is the Debye function.

{\it{(iii) Mixed coupling scheme:}} 
The predictions of DDFTs based on local or non-local schemes have been compared
to simulations, and both were found to have shortcomings\cite{Reister2003,
Qi2017}. In a previous paper\cite{Qi2017}, two of us have therefore proposed a
mixed scheme where the dynamics is assumed to be governed by a local mobility
function on short wavelengths and a nonlocal one on large wavelengths. To
this end, a filter function $\Gamma\left(\rr\right)$ was introduced that
filters out the long-wavelength part of the thermodynamic driving force via a
convolution integral
\begin{equation}
\hat{\ff}^{\mbox{\tiny Nonlocal}}_\alpha\left(\rr\right) 
  = - \int \ud\rr' \Gamma\left(|\rr-\rr'|\right) \nabla \hmu_\alpha(\rr').
\label{eq:MixGlobal}
\end{equation}
with
\begin{equation}
\Gamma\left(r\right) = \left(2\pi\sigma^2\right)^{-3/2} \exp\{-r^2/2\sigma^2\}.
\label{eq:MixFilter}
\end{equation}
This ''coarsened'' force is then taken to drive nonlocal chain diffusion,
whereas the remaining part, 
\begin{equation}
\hat{\ff}^{\mbox{\tiny Local}}_\alpha(\rr) = - \nabla \hmu_\alpha(\rr)
 - \hat{\ff}^{\mbox{\tiny Nonlocal}}_\alpha(\rr)
\label{eq:MixLocal}
\end{equation}
drives local rearrangements of the chain via a local mobility coefficient.
The resulting interpolated scheme has the form
\begin{eqnarray}
\frac{\partial \phi_\alpha}{\partial t} 
  & =& -\nabla \sum_\beta \int \ud \rr' \left[ 
      \hL^{\mbox{\tiny Nonlocal}}_{\alpha\beta}\left(\rr,\rr'\right)
     \hat{\ff}^{\mbox{\tiny Nonlocal}}_\beta\left(\rr'\right) \right.
\nonumber \\ & & \left. \qquad \quad \qquad
     + \: \hL^{\mbox{\tiny Local}}_{\alpha\beta}\left(\rr,\rr'\right)
     \hat{\ff}^{\mbox{\tiny Local}}_\beta\left(\rr'\right)\right],
\label{eq:MixDDF}
\end{eqnarray}
where $\hL^{\mbox{\tiny Nonlocal}}$ can be any of the chain coupling schemes
discussed above. The tunable parameter $\sigma$ determines the length scale of
crossover between the local and the nonlocal dynamics. When referring to mixed
scheme DDFT calculations in the present paper, these are carried out by mixing
local and Debye dynamics with the filter parameter $\sigma=0.3 \Rg$, a value
found to be optimal in our previous work\cite{Qi2017}. 

\section{Approaches to determine DDFT mobility coefficients
from microscopic simulations}

The expressions for the mobility coefficients discussed in the previous section
were postulated more or less heuristically, without much input on the
underlying microscopic dynamics. The only parameters that can be used to match
the microscopic and the DDFT dynamics are the diffusion constant, and in case
of the mixed scheme, the tuning parameter $\sigma$. The purpose of the present
work is to derive more informed bottom-up schemes, where the mobility
coefficients are calculated from simulations of a microscopic reference system.
We have explored two such approaches which we will now discuss below.

In both cases, we will assume that our system is homogeneous, hence
$\Lambda(\rr,\rr')´$ is translationally invariant. We can then conveniently
rewrite the DDFT equations in Fourier representation as
\begin{equation}
\partial_t \rho_\alpha(\qq,t) = - q^2 \sum_\beta 
  \Lambda_{\alpha \beta}(\qq) \mu_\beta(\qq,t)
\label{eq:DDFT_f}
\end{equation}
with $\mu_\beta(\qq,t)/V = \delta F/\delta \rho_\beta(-\qq,t)$. 
Here and throughout, we define the Fourier transform via \cite{Doi} 
\begin{displaymath}
f({\qq}) = \int d \rr e^{i \qq \cdot \rr} f(\rr), \quad
f(\rr) = \frac{1}{V} \sum_{\qq} e^{- i \qq \cdot \rr} f(\qq).
\end{displaymath}

\subsection{Green-Kubo approach}

\newcommand{\LGK}{\Lambda^{(s),\mbox{\tiny GK}}}

The first approach is based on the Green-Kubo formalism, which is a standard
tool to determine transport coefficients from simulations.  Let us first
recapitulate the general formalism.\cite{Zwanzig,Hansen,Kubo1996} For a given
microscopic system with Hamiltonian $H$, we consider the linear response of a
quantity $\dot{A}$ to a perturbation of $H$ caused by a generalized field $Z_B$
that couples to a quantity $B$ (i.e., $H = H_0 - Z_B B$).  According to the
Green-Kubo formalism, the response is given by $\langle \dot{A} \rangle =
\lambda_{AB} Z_B$ with \mbox{$\lambda_{AB} = - \frac{1}{k_B T} \int_0^{\infty}
\ud t \: \langle \dot{A}(t) \dot{B}(0) \rangle$} in classical systems.

To apply this formalism to our DDFT problem, we choose $A = \rho_\alpha(\qq,t)$
and $B = \rho_\beta(\qq,t)$, where $\rho_\zeta(\qq,t)$ (with $\zeta =
\alpha,\beta$) is derived from the monomer coordinates $\RR_k(t)$ via
$\rho_\zeta(\qq,t) = \sum_k \: e^{i \qq \cdot \RR_k(t)} \gamma_k^{(\zeta)}$
with $\gamma_k^{(\zeta)}=1$ if monomer $k$ is of type $\zeta$, and
$\gamma_k^{(\zeta)}=0$ otherwise. This results in \mbox{$\dot{A} = i \qq \cdot
\jj_\alpha(\qq,t)$} and \mbox{$\dot{B} = - i \qq \cdot \jj_\beta(\qq,t)$} with
$\jj_\zeta(\qq,t) = \sum_k \: e^{i \qq \cdot \RR_k(t)} \dot{\RR}_k(t)
\gamma_k^{(\zeta)}$.  The continuity equation for $\rho_\alpha$ in Fourier
representation reads \mbox{$\partial_t \rho_\alpha(\qq,t) = i \qq \cdot
\jj_\alpha(\qq,t) = \dot{A}$}. From Eq.\ (\ref{eq:DDFT_f}), we hence know
$\dot{A} = - q^2 \sum_\beta \Lambda_{\alpha \beta}(\qq) \mu_\beta(\qq)$, where
$(- \mu_\beta(\qq,t)/V)$ couples to $B$.  Now, in the linear response regime,
an external field $Z_B$ coupling to $B$ would contribute additively to $(-
\mu_\beta(\qq,t)/V)$ and generate the same response, hence we can identify
$\Lambda_{\alpha \beta} = \lambda_{AB}/q^2 V$ and  the Green-Kubo formalism
results in the following expression: 
\begin{equation} 
\label{eq:GK}
\Lambda_{\alpha \beta}(\qq) 
    = \frac{1}{V k_B T} \int_0^\infty \!\!\! \ud t \:
      \Big\langle \jj_\alpha(\qq,t) \jj_\beta(-\qq,0) \Big\rangle : \hq \hq,
\end{equation} 
with $\hq = \qq/q$ and the tensor products $\jj \jj$ and $\hq \hq$.

\revision{However, the numerical evaluation of this expression and a
theoretical analysis for the special case of Rouse chains shows that
Eq.\ (\ref{eq:GK}) yields zero for all nonzero $\qq$. This is
demonstrated in more detail in the appendix. Only at $\qq = 0$ do we
recover the familiar Green-Kubo expression for the diffusion constant.
}

The reason becomes clear if we recall the premises underlying the Green-Kubo
relations. They describe the response of {\em stationary} currents to
generalized thermodynamic forces. In our case, at $q \neq 0$, a stationary
current is not possible, since it would generate indefinitely growing density
fluctuations $\rho(\qq,t)$. Since $\rho(\qq,t)$ must saturate eventually, the
flows $\jj(\qq,t)$ will average to zero at late times, independent of the
applied generalized forces. Therefore, the Green-Kubo transport coefficients
must vanish for any nonzero $\qq$.  Stationary currents are only possible at
$\qq=0$. Hence the Green-Kubo formalism is not suitable for determining
$\qq$-dependent mobility functions for DDFT models.

In fact, this problem is not uncommon in applications of Green-Kubo integrals
\cite{Kirkwood1949,Pep2019}. For example, confinement can prevent stationary
currents, which is why Green-Kubo integrals may vanish in confined systems,
even if locally, a description in terms of a Markovian dynamical equations with
well-defined transport coefficients is appropriate.  The $\qq$-dependent
Green-Kubo integrals considered here, which describe the response to a
spatially varying field, vanish for a similar reason.  One popular solution to
this problem has been to assume that the time scales of local Markovian
dynamics and global constrained dynamics are well separated, and to search for
a plateau in the running Green-Kubo integrals. In our case, however, the
running integrals do not exhibit a well-defined plateau (data not shown).  We
will discuss this point further in Sec.\ \ref{sec:discussion}

\subsection{Relaxation time approach}

\label{sec:relaxation}

In the present subsection, we describe an alternative approach to deriving DDFT
mobility coefficients from microscopic trajectories: We propose to estimate
them directly from the characteristic relaxation time of the single chain
dynamic structure factor.

To motivate our Ansatz, we begin with discussing some implications of the DDFT
equations.  We consider the dynamics of a single tagged chain $s$ with
corresponding monomer density $\rho_{\alpha}^{(s)}$. In the mean-field spirit,
the DDFT equation for $\rho_{\alpha}^{(s)}$ in Fourier representation takes the
form
\begin{equation}
\label{eq:drhos}
\partial_t \rho_{\alpha}^{(s)}(\qq,t) = - q^2  \sum_\beta
\Lambda_{\alpha \beta}^{(s)} (\qq) \: \mu_{\beta}^{(s)}(\qq, t),
\end{equation}
where $\Lambda^{(s)} = \hL \: N^2$ is the mobility per chain,
and $\mu_{\beta}^{(s)}(\qq) = V \delta F^{(s)}/\delta \rho_{\beta}^{(s)}(-\qq)$
is derived from the free energy $F^{(s)}$ of a single chain that moves
independently in the averaged background provided by the other chains.
Next we multiply both sides with $\rho_{\gamma}^{(s)}(-\qq, 0)$ and
\revision{average over chain conformations}. Identifying
$g_{\alpha \gamma}(\qq, t) = \frac{1}{N} 
\langle \rho_{\alpha}^{(s)}(\qq,t) \rho_{\gamma}^{(s)}(-\qq,0) 
\rangle$,  
we obtain
\begin{equation}
\partial_t g_{\alpha,\gamma}(\qq,t) = - \frac{q^2}{N}  \sum_\beta
\Lambda_{\alpha \beta}^{(s)} (\qq) 
\left< \mu_{\beta}^{(s)}(\qq, t) \rho_{\gamma}^{(s)}(- \qq, 0) \right>.
\label{eq:dgt}
\end{equation}
To proceed, we expand $F^{(s)}$ in powers of $\rho^{(s)}(\qq)$, giving
\begin{equation}
\label{eq:fs}
F^{(s)} = \mbox{const.} +  \frac{k_B T}{2NV} \sum_{\qq}
\vecR^{(s)}(-\qq) \matG^{-1}(\qq,0) \vecR^{(s)}(\qq) + \cdots
\end{equation}
Here and in the following, we use a matrix notation for convenience,
i.e. $\vecR \corresponds ( \rho_\alpha )$,
$\matL \corresponds ( \Lambda_{\alpha \beta} )$
etc. Taking the derivative with respect to $\rho_{\beta}^{(s)}(- \qq)$,
we obtain \mbox{$ \vecM^{(s)}(\qq) \approx k_B T \frac{1}{N} 
\matG^{-1}(\qq) \vecR^{(s)}(\qq) $}.
Inserting this in Eq.\ (\ref{eq:dgt}) yields
\begin{equation}
\partial_t \matG(\qq,t) \approx - \frac{k_B T q^ 2}{N}
\matL^{(s)}(\qq) \: \matG^{-1}(\qq,0) \: \matG(\qq,t),
\label{eq:g_DDFT}
\end{equation}
which can be solved in matrix form, giving
\begin{equation}
\matG(\qq,t) = \exp \left(- \frac{k_B T q^2}{N} 
\matL^{(s)}(\qq) \matG^{-1}(\qq,0) t \right) \matG(\qq,0).
\end{equation}
This equation approximates the relaxation of the single chain under
three assumptions: (i) Memory effects were neglected (the basis 
of the DDFT approach), (ii) a mean-field approximation was made 
(in Eq.\ (\ref{eq:drhos})), and (iii) density fluctuations were
assumed to be small (in Eq.\ (\ref{eq:fs})). Within these approximations,
the relaxation of the chain is determined by a $\qq$-dependent
''relaxation time matrix'' $\matT(\qq)$, 
$\matG(\qq,t) = \exp(-t \:\matT^{-1}(\qq)) \: \matG(\qq,0)$
and, using $\matL^{(s)} = \hmatL\: N^2$, we can identify
\begin{equation}
\hmatL(\qq) = \frac{1}{k_B T q^2 N}\: \matT^{-1}(\qq)  \: \matG(\qq,0). 
\label{eq:lambda1}
\end{equation}
We can further simplify this expression by assuming that the relaxation of the
chain is governed by a single $\qq$-dependent time constant $\tau(\qq)$, i.e.,
$\matT(\qq) \approx \mathbb{1} \cdot \tau(\qq)$.  Then Eq.\
(\ref{eq:lambda1}) can be rewritten as 
\begin{equation}
\hmatL(\qq) = \frac{1}{k_B T q^2 N \: \tau(\qq) }  \: \matG(\qq,0). 
\label{eq:lambda2}
\end{equation}

The considerations above suggest the following procedure to determine an
effective mobility coefficient for the DDFT model: We first conduct
fine-grained simulations of the polymer melt in a homogeneous reference system
(i.e., in the case of the diblock copolymer melt, below the order-disorder
transition (ODT)).  From the simulation trajectory for the full $g(\qq,t)$, we
compute the relaxation time $\tau(\qq)$ and insert it in the expression
(\ref{eq:lambda1}) or (\ref{eq:lambda2}). 

The question remains how to define the characteristic relaxation time. This 
question is non-trivial, because the actual behavior of $g(\qq,t)$ is driven by
a multitude of time scales, corresponding to the different internal modes of
the chain. At late times, the slowest diffusive mode dominates, and $g(\qq,t)$
has the limiting behavior \cite{Doi} $\lim_{t \to \infty} g(\qq,t) \propto
\exp(- D_c q^2 t)$, giving $\tau = 1/D_c q^2$. Inserting this in
(\ref{eq:lambda2}), we recover the Ansatz of nonlocal Debye dynamics, (see
(\ref{eq:Debye})) $\hmatL(\qq) = \frac{D_c}{N k_B T} \: \matG(\qq)$. 

However, by the time this limiting behavior sets in, much of the
structuring has already taken place.  It would be more desirable to define
$\tau(\qq)$ such that it captures the dominant time scales of structure
formation on the scale $\qq$. In the present work, we test two prescriptions
for determining $\tau$ and then calculate $\hmatL$ via Eq.\ (\ref{eq:lambda2}):
\begin{eqnarray}
\label{eq:tauR}
\hmatL^{\tau_R}: \quad \mbox{from} \quad &&
\tau_R = \frac{1}{g(\qq,0)} \int_0^\infty \ud t \: g(\qq,t),
\\
\label{eq:tauE}
\hmatL^{\tau_e}: \quad \mbox{from} \quad &&
g(\qq, t=\tau_e) \stackrel{!}{=} g(\qq,0)/e,
\end{eqnarray}
where \revision{$e$ is the Euler number and} $g(\qq,t)$ is the full 
single-chain structure factor,
\begin{equation}
\label{eq:gfull}
g(\qq,t) = \sum_{\alpha, \beta} g_{\alpha \beta}(\qq, t).
\end{equation}
In a third approach, we generalize (\ref{eq:tauR}) to extract
a full relaxation time matrix,
\begin{equation}
\label{eq:tauFull}
\hmatL^{T}: \quad \mbox{from} \quad 
\matT(\qq) = \int_0^\infty \ud t \: \matG(\qq,t) \: \matG^{-1}(\qq,0).
\end{equation}
and use that to determine $\hmatL$ via Eq.\  (\ref{eq:lambda1}).
Calculating $\hmatL$ with this method involves matrix inversions and
multiplications for every value of $\qq$. However, in the case of {\em
symmetric} A:B diblock copolymers with fully equivalent $A$ and $B$ blocks, the
prescription can be simplified.  For symmetry reasons, $\matG$, $\matT$ and
$\hmatL$ then have the same matrix structure $(M_{\alpha \beta})$
with $M_{AA} = M_{BB}, M_{AB} = M_{BA}$ and thus share the same 
Eigenvectors, $(1,1)$ and $(1,-1)$.  
Using these to diagonalize $\matG$ and $\matT$, we obtain 
\begin{eqnarray}
\hL_{AA}(\qq) &=& \frac{1}{4 k_B T q^2 N} 
  \left( \frac{g(\qq,0)}{\tau_R} + \frac{\Delta(\qq, 0)}{\tau_\Delta} \right)
\\
\hL_{AB}(\qq) &=& \frac{1}{4 k_B T q^2 N} 
  \left( \frac{g(\qq,0)}{\tau_R} - \frac{\Delta(\qq, 0)}{\tau_\Delta} \right)
\end{eqnarray}
with $g(\qq,t)$ and $\tau_R$ defined as above 
(Eqs.\ (\ref{eq:gfull}), (\ref{eq:tauR})),
$\Delta(\qq,t) = g_{AA}(\qq,t) + g_{BB}(\qq,t)
  - g_{AB}(\qq,t) - g_{BA}(\qq,t)$, and
$\tau_\Delta = \frac{1}{\Delta(\qq,0)} \int_0^\infty \ud t \: \Delta(\qq,t)$.

In practice, determining the integrals (\ref{eq:tauR}) and (\ref{eq:tauFull})
by numerical integration of simulation data only is not possible for small $q$,
because the relaxation time diverges for $q \to 0$. Therefore, an extrapolation
procedure must be devised. At late times, $g_{\alpha \beta}(\qq,t)$ is known to
decay exponentially\cite{Doi} according to $g(\qq,t) \sim \exp(-q^2 D_c t)$. 
Hence we make the Ansatz
 \begin{equation}
 g_{\alpha\beta}(q,t) = g_{\alpha\beta}\left(q,t_i)\right)
    \exp\left(-q^2 D_{\mbox{\tiny eff}} \left(t-t_i\right)\right), 
 \label{eq:gLT}
 \end{equation}
for large $t, t_i$ with $t>t_i$. Specifically, we fit the data for
$g_{\alpha\beta}\left(q,t\right)$ to Eq.\ (\ref{eq:gLT}) in time
windows $t \in [t_i, t_f]$, using the weighted least squares fit module
in the Matlab suite \cite{Matlab}, and then choose those values of $t_{i,f}$
which yield the value of $D_{\mbox{\tiny eff}}$ that is closest to the
theoretical value, $D_c = D_0/N$. The integrals over $t$ in 
(\ref{eq:tauR}) and (\ref{eq:tauFull}) are then evaluated by first
numerically integrating the data up to $t = t_i$, and then using
the extrapolation (\ref{eq:gLT}) in the integral from $t=t_i$ to infinity.
Typical values for $t_i, t_f$ are $t_i \approx 20 t_0$ and 
$t_f \approx 40 t_0$, where $t_0$ is the simulation time unit, see
below.

\begin{figure}[t]
\includegraphics[scale=0.32]{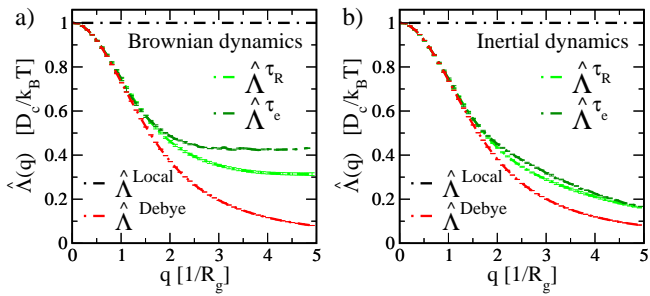}
\caption{Normalized mobility functions of homopolymers with length $N=40$ in a
melt, as obtained {\em via} the relaxation time method (Eq.\
(\ref{eq:lambda2})) with data from Brownian dynamics (a) and inertial dynamics
(b) simulations. Two prescriptions for determining the single chain relaxation
time are tested, $\tau_R$ (green, Eq.\ (\ref{eq:tauR})) and $\tau_e$ (blue,
Eq.\ (\ref{eq:tauE})). Also shown for comparison are the results from the Debye
and the local approximation (($\Lambda^{\mbox{\tiny Debye}}\left(q\right)$,
red) and ($\Lambda^{\mbox{\tiny Local}}\left(q\right)$, black)).}
\label{fig:2}
\end{figure}

Fig.\ \ref{fig:2} shows results for the $q$-dependent mobility functions of
homopolymers in a homopolymer melt. They were extracted from Brownian dynamics
simulations (massless monomers, Fig.\ \ref{fig:2}a) and molecular dynamics
simulations (massive monomers, Fig.\ \ref{fig:2}b) of melts of Gaussian chains
with length $N=40$, using the prescriptions (\ref{eq:tauR}) and
(\ref{eq:tauE}). We note that in the case of homopolymers, the prescription
(\ref{eq:tauFull}) is equivalent to (\ref{eq:tauR}). For comparison, we also
show the mobility functions corresponding to the local and the Debye
approximation. In the local scheme, the mobility is constant, in the Debye
scheme, it is proportional to the static structure factor. The results from the
relaxation schemes are intermediate between the local and the Debye scheme.  At
small $q$, they follow the Debye scheme. At larger $q$, the mobility
is enhanced, hence small wavelength modes relax faster. The effect is more
pronounced for Brownian dynamics than for inertial dynamics, most likely
because the inertial time scale contributes to the total relaxation time at
small wavelengths (see also Fig. \ref{fig:1} b).

Thus we find that the mobility functions obtained with the relaxation time
approach interpolate between the nonlocal mobility function (at small $q$) and
the local mobility function (at larger $q$). This seems promising, since our
previous studies have suggested that such an interpolation may be necessary to
capture the kinetics of structure formation in copolymer systems\cite{Qi2017}.
We will now test our DDFT approach by performing a systematic comparison of
fine-grained simulations and DDFT predictions for the ordering/disordering
kinetics in block copolymer melts.

\section{Application to diblock copolymer melts}

We consider melts of $n_c$ block copolymers containing $N_A$ beads of type $A$
and $N_B$ beads of type $B$, in a box of volume $V=L_x\times L_y\times L_z$
with dimension $L_i$ in $i$ direction and periodic boundary conditions.  The
average monomer density is thus $\rho_0 = n_c N/V$.  Polymers are modelled as
Gaussian chains, i.e., chains of ''monomer beads'' connected by harmonic
springs. The non-bonded monomer interactions are characterized in terms of a
Flory Huggins parameter $\chi$, which controls the incompatibility between $A$
and $B$ monomers, and a Helfand parameter $\kappa$, which controls the
compressibility.  

We carry out fine-grained simulations of order/disorder processes in such
systems and compare them with DDFT calculations, using the SCF free energy
functional and mobility functions that are extracted from fine-grained
simulations at $\chi=0$.  

Throughout this paper, lengths will be represented in units of the radius of
gyration $\Rg$ of an ideal chain of length $N=N_A+N_B$, energies in units of
the thermal energy, $k_B T$, and time in units of $t_0 = R_g^2/D_0$, where
$D_0$ is the monomer diffusivity.  

\subsection{Model and methods}

\subsubsection{Fine-grained model and simulation method}

\label{sec:simulations}

Since we focus on a comparison of dynamical properties of particle-based and
field-based models here, we use as fine-grained model a particle-based
implementation of an Edwards model\cite{Edwards1965,Laradji1994,Ganesan2003,
Daoulas2006}, where the non-bonded monomer interactions are described by the
same Hamiltonian than that underlying the SCF free energy functional. At
sufficiently high polymer density and sufficiently far from critical points,
the static properties of such models are known to be well represented by SCF
functionals without much parameter adjustment \cite{Qi2017}. 

Non-bonded interactions are thus expressed as a functional of the local monomer
densities\cite{Laradji1994}. Let $\RR_{m,j}$ denote the position of the $j$th
monomer on the $m$th chain. The Hamiltonian $H$ describing the monomer
interactions is then expressed as
\begin{eqnarray}
\nonumber
H/k_B T &= &\frac{N}{4\Rg^2}\sum_{m=1}^{n_c}\sum_{j=1}^{N}
        \left(\RR_{m,j}-\RR_{m,j-1}\right)^2 
\\ &&
\nonumber
   + \rho_0 \chi \int \ud \rr \:
          \hat{\phi}_A\left(\rr\right)\hat{\phi}_B\left(\rr\right) 
\\&&
   + \rho_0 \kappa \int \ud \rr 
    \left(\hat{\phi}_A\left(\rr\right) 
           + \hat{\phi}_B\left(\rr\right) -1 \right)^2,
\label{eq:H}
\end{eqnarray}
where the first term represents the bonded interactions in the polymer, and the
last two terms correspond to non-bonded interactions.  The quantities
$\hat{\phi}_\alpha\left(\rr\right)$ are the normalized microscopic densities of
$\alpha$-type beads ($\alpha=A$ or $B$) at position $\rr$, defined as,
$\hat{\phi}_\alpha\left(\rr\right) =
\frac{1}{\rho_0}\sum_{mj}\delta\left(\rr-\RR_{mj}\right)
\delta_{\alpha,\tau_{mj}}$, where $\tau_{mj}=A$ or $B$ characterizes the
monomer sequence on chain $m$.  

In practice, the local densities are evaluated on a grid with grid
size $\Delta x = \Delta y = \Delta z = 0.1 \Rg$, using a first order
cloud in the cell (CIC) scheme\cite{Birdsall1969}. The grid size is an
important ingredient of the model definition, as it sets the range of
non-bonded interactions.  In the simulations, we consider systems with
average monomer density \mbox{$\rho_0 = \frac{2}{3} \cdot 10^5
/R_g^3$} , i.e., roughly 50 monomers per grid cell. For this choice of
densities and grid parameters, grid
artefacts\cite{Detcheverry2008,Qi2015} \revision{are negligible, and the
renormalized values of $\chi$ and $\kappa$ in the SCF theory are
practically identical to the corresponding ''bare'' parameters in 
Eq.\ (\ref{eq:H}) \cite{Qi2015}). Furthermore, fluctuation effects
are small.} The strength of thermal fluctuations can be characterized by
the Ginzburg parameter\cite{Fredrickson,Muller2005}, $C^{-1} = V/n_c \Rg^3$.
In our system, this parameter is $C^{-1}=0.01$ or less.

Monomers $(m,j)$ with mass $M_{m,j}$ evolve in time according to a 
Langevin equation,
\begin{equation}
  M_{m,j} \dot{\vv}_{m,j}\left(t\right)
    = - \frac{\partial H}{\partial \RR_{m,j}}  
      - \Gamma \vv_{m,j} + \sqrt{2\Gamma k_B T} \; \ff_{m,j}\left(t\right).
\label{eq:ID}
\end{equation}
The first term on the right hand side describes the conservative
interaction forces, the second term corresponds to a friction force (with $\vv
= d\RR/dt$ and the monomer friction $\Gamma = 1/D_0$), and the last term to a
stochastic force representing the effect of thermal fluctuations, where
$\ff_{mj}\left(t\right)$ is a Gaussian distributed random noise with zero mean
and variance $\left<\ff_{mj}\left(t\right)\ff_{nk}\left(t'\right)\right>
=\delta_{mn}\delta_{jk}\mathbb{1}\delta\left(t-t'\right)$.  Hydrodynamic
interactions are thus neglected, and since the interaction potentials defined
by Eq.\ (\ref{eq:H}) are soft, entanglement effects are not included as well.
We consider the two cases $M_{m,j} \equiv 1 k_BT t_0^2/\Rg^2$ (inertial
dynamics), and $M_{m,j} \to 0$ (overdamped dynamics). In the second case, Eq.\
(\ref{eq:ID}) is replaced by 
\begin{equation} 
   \frac{d\RR_{m,j}}{dt} =
          -D_0\frac{\partial H}{\partial \RR_{m,j}} 
          +\sqrt{2D_0 k_B T}\; \ff_{m,j}\left(t\right).  
\label{eq:BD} 
\end{equation} 
The equations of motion are integrated using the Velocity-Verlet
scheme\cite{Frenkel,Brunger1984} in the case of inertial dynamics (Eq.\
(\ref{eq:ID})), and the Euler-Maruyama\cite{Frenkel} algorithm in the case of
overdamped dynamics (Eq.\ (\ref{eq:BD})) with the time step $\delta t = 0.001
t_0$. 

Specifically, we consider copolymer melts in a simulation box of size $\Rg
\times \Rg \times 3 \Rg$.  Unless stated otherwise, we consider symmetric
copolymers, i.e., $N_A = N_B$, with total length $N=40$. For comparison, we
also study copolymers with length $N=20$ or $N=100$, and vary the A:B fraction.
In all cases the monomer density is kept fixed at \mbox{$\rho_0 =
\frac{2}{3} \cdot 10^5  /R_g^3$}.  The Helfand parameter is set to $\kappa N =
100$.  The systems are initially prepared by growing polymers at randomly
picked points in the simulation box. In three independent runs, configurations
are then equilibrated for 300000 time steps each. Data for $g(\qq,t)$ are
subsequently collected over 200000 time steps and used to extract the mobility
functions. 
In a set of additional simulations, we monitor the formation of lamellar
structure in the melt after a step change from $\chi N = 0$ to a finite $\chi
N$ above the ODT, and the decay of the lamellar structure after a step change
from finite $\chi N$ to $\chi N = 0$.  The systems are equilibrated as
described above and the time evolution is then monitored over 100000 time steps
in 10 independent runs. 

\subsubsection{SCF free energy functional}

As discussed earlier, we use the SCF theory to construct the free energy
functional in our DDFT equations. The SCF theory is one of the most powerful
equilibrium theories for inhomogeneous polymer systems and has been well
documented elsewhere \cite{Schmid1998,Matsen2002,Fredrickson, Schmid2011}.
Here, we just briefly summarize the main equations, adjusted to our system. We
model the copolymers as continuous Gaussian chains
\cite{Matsen2002,Schmid2011}, and parameterize the contour length by a
continuous variable $s \in [0:1]$.  The free energy functional
$F\left[\{\phi_\alpha\left(\rr\right)\}\right]$ of our block copolymer system
is expressed as
\begin{eqnarray}
\label{eq:FE}
F/k_B T
&=& \frac{\rho_0}{N} 
  \bigg\{ \int \ud \rr \Big[
     \chi N \: \phi_A\left(\rr\right)\phi_B\left(\rr\right) 
\\ && \qquad
   + \kappa N\left(\phi_A\left(\rr\right) + \phi_B\left(r\right)-1\right)^2
      \Big]
\nonumber \\ &&  \nonumber
 - \sum_{\alpha=A,B} \int \ud \rr \: \phi_\alpha\left(r\right)
   \omega_\alpha\left(\rr\right) -V\ln Q\bigg\},
\end{eqnarray}
where $\phi_\alpha$ is the normalized density field of monomers
of type $\alpha$, $\omega_\alpha$ the corresponding conjugate field,
and $Q$ is the single chain partition function in the external field 
$\omega_\alpha$.  The conjugate fields are determined implicitly by
the requirement
\begin{eqnarray}
\phi_A\left(\rr\right) 
   &=& \frac{V}{Q}\int_0^{N_A/N} \ud s  \:
       q_f\left(\rr,s\right) q_b\left(\rr,1-s\right),
   \nonumber \\
  \phi_B\left(\rr\right) 
   &=& \frac{V}{Q}\int_0^{N_B/N} \ud s \:
       q_b\left(\rr,s\right) q_f\left(\rr,1-s\right).
\label{eq:SCF1}
\end{eqnarray}
Here $q_f\left(\rr,s\right)$ and $q_b\left(\rr,s\right)$ are the 
end-integrated forward and backward chain propagators, respectively,
which can be obtained from solving the following differential equation: 
\begin{equation}
\frac{\partial q\left(\rr,s\right)}{\partial s} 
  = \Rg^2 \nabla^2q\left(\rr,s\right) 
    - \omega\left(\rr\right) q\left(\rr,s\right)
\label{eq:Green}
\end{equation}
with initial condition $q_{f,b}\left(\rr,0\right)=1$ and
\mbox{$\omega\left(\rr\right) = \omega_A\left(\rr\right)$} or
$\omega_B\left(\rr\right)$, depending on $s$: 
$q_f\left(\rr,s\right)$ is obtained by setting
$\omega\left(\rr\right)=\omega_A\left(\rr\right)$ for $s<N_A/N$ and
$\omega\left(\rr\right)=\omega_B\left(\rr\right)$ otherwise, and
$q_b\left(\rr,s\right)$ by setting
$\omega\left(\rr\right)=\omega_B\left(\rr\right)$ for $s<N_B/N$ and
$\omega\left(\rr\right)=\omega_A\left(\rr\right)$ otherwise.
Knowing $q_f$ or $q_b$, one can calculate the single chain partition
function $Q$ via
\begin{equation}
  Q = \frac{1}{V}\int \ud \rr \: q_f\left(\rr,1\right) 
    = \frac{1}{V}\int \ud \rr \: q_b\left(\rr,1\right)
\label{eq:PartFn}
\end{equation}

At equilibrium, $F\left[\{\phi_\alpha\left(\rr\right)\}\right]$ 
assumes a minimum with respect to $\phi_\alpha\left(\rr\right)$,
leading to a second set of conditions for the values of the 
conjugate fields,
$\omega_\alpha$:
\begin{eqnarray}
 \omega_A^{\mbox{\tiny SCF}}\left(\rr\right) 
    & =& \chi N\phi_B + 2\kappa N\left(\phi_A + \phi_B -1\right)
 \nonumber \\
  \omega_B^{\mbox{\tiny SCF}}\left(\rr\right) 
     & = &\chi N\phi_A + 2\kappa N\left(\phi_A + \phi_B -1\right)
\label{eq:SCF2}
\end{eqnarray}

However, in DDFT calculations, these conditions are not imposed. Instead,
the system is dynamically driven towards the equilibrium state {\em via}
the diffusive dynamical equation (\ref{eq:DDFT}) with
$\hmu_\alpha(\rr)=(\omega_\alpha^{\mbox{\tiny SCF}}(\rr)-\omega_\alpha(\rr))$.

The SCF and DDFT calculations in the present work are effectively
one-dimensional, i.e., we assume that densities vary only in the $z$ direction.
Space is discretized with grid size $\Delta z = 0.1\Rg$ The propagator
equation, Eq.\ (\ref{eq:Green}) is solved using the pseudo spectral scheme
\cite{Fredrickson} with discretization $\Delta s = 0.01$.  As in our earlier
work\cite{Qi2017}, the time step in the DDFT calculations depends on the DDFT
scheme: We use $\Delta t = 10^{-4}t_0 N $ for DDFT calculations based on Debye
dynamics or any of the other pre-determined mobility functions $\hL(\rr -
\rr')$ discussed in Sec.\ \ref{sec:relaxation}, $\Delta t = 10^{-5}t_0 N $ for
full chain dynamics, Eq.\ (\ref{eq:FullChain}), and $\Delta t = 10^{-6}t_0 N $
for local dynamics (\ref{eq:Local}) or mixed dynamics (\ref{eq:MixDDF}).

\subsection{Mobility functions}

Based on simulations of the fine-grained model discussed above, mobility
functions were extracted from the simulation data using the different variants
of the relaxation time approaches discussed in Section \ref{sec:relaxation}.
In the following, we will consider melts of symmetric A:B diblock copolymer
melts.

\begin{figure}[t]
\includegraphics[scale=0.32]{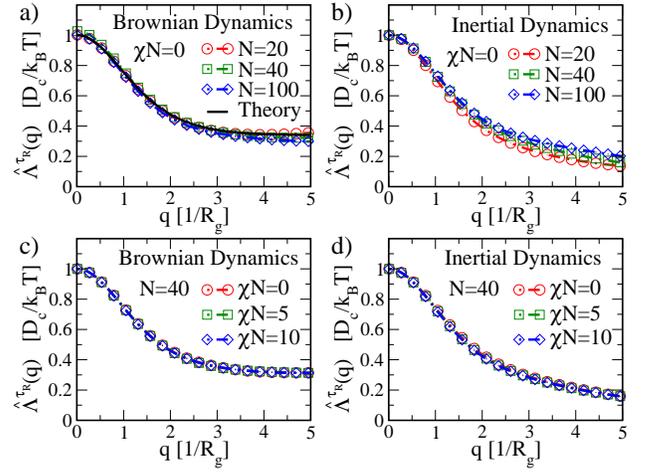}
\caption{
Normalized full-chain mobility function
of symmetric A:B copolymers in a melt, as obtained {\em via} the
relaxation time method (Eq. (\ref{eq:tauR})) from Brownian dynamics
simulations (a,c) and inertial dynamics simulation data (b,d) for different
chain lengths $N$ or interaction parameter $\chi$ as indicated.
Solid line in (a) shows theoretical prediction for $N=40$
obtained by inserting
Eq.\ (\ref{eq:Doi}) into Eq.\ (\ref{eq:tauR}).  }
\label{fig:3}
\end{figure}

Fig.\ \ref{fig:3} shows the results for the full-chain mobility function, $\hL
(q)= \sum_{\alpha \beta} \hL_{\alpha \beta}(q)$  for different chain lengths
($N=20,40,100$) at fixed $\chi = 0$, and for different values of $\chi N$
($\chi N = 0,5,10$) at fixed chain length $N=40$. These values were chosen such
that $(\chi N)$ is still below the value \cite{Gehlsen1992, Rosedale1995} $(\chi
N)_{ODT}\approx10.5$ where the order-disorder transition sets in for symmetric
diblock copolymers, hence the melt is disordered and isotropic.  The
behavior of $\hL(q)$ at $q \to 0$ reflects the translational diffusion of
chains and takes the asymptotic value $\hL = D_c$. Therefore, the curves are
rescaled by the chain diffusion constant $D_c$, which has been calculated
independently from the mean-square displacement of the chain.  For example, for
$N=40$, we obtain $D^{BD}_c=(0.0263\pm0.0001)\Rg^2/t_0$ in Brownian dynamics
simulations, and $D^{ID}_c=(0.0224\pm0.0003) \Rg^2/t_0$ in inertial dynamics
simulations, which is close to the value for free Rouse chains, $D_c=0.025
\Rg^2/t_0$. Since the interactions between monomers are very soft in the
particle-based model, they do not affect the diffusion constant significantly
in the disordered phase. 

The full-chain mobility function is found to depend weakly on the chain length
$N$ (Fig.\ \ref{fig:3}a,b), the effects being most pronounced in the
regime of high $q$: If one increases $N$, the mobility function for high $q$
decreases in the Brownian dynamics case and increases in the inertial dynamics
case, such that both mobility functions approach each other. In contrast, the
Flory Huggins parameter $\chi$ has practically no influence on the chain
mobility function in the disordered regime (Fig.\ \ref{fig:3}c,d)). Motivated
by this finding, we will use the mobility functions obtained at $\chi=0$ in all
DDFT calculations below.  

\begin{figure}[t]
\includegraphics[scale=0.32]{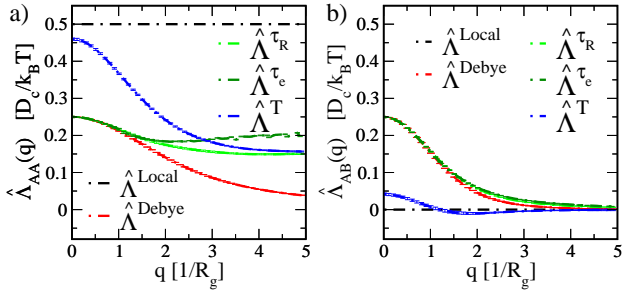}
\caption{
Normalized mobility function $\hL_{\alpha \beta}$ of
symmetric A:B diblock copolymers (length $N=40$), obtained from
Brownian dynamics simulations at $\chi = 0$,  using different variants
of the relaxation time method: Eqs.\ (\ref{eq:tauR}) (light green line), 
(\ref{eq:tauE}) (dark green line), and (\ref{eq:tauFull}) (blue line).
Also shown for comparision is the result from the Debye approximation 
(red line) and the local approximation in a homogeneous melt (black).
}
\label{fig:4}
\end{figure}

Next we turn to the discussion of the monomer-species resolved mobility
functions $\hL_{\alpha \beta}$. The results extracted from Brownian dynamics
simulation trajectories for symmetric diblock copolymers of length $N=40$ are
shown in Fig.\ \ref{fig:4} for the different relaxation time approaches
discussed in Sec.\ \ref{sec:relaxation}. Since $\hL_{AA}(q)=\hL_{BB}(q)$ for
symmetric systems, and $\hL_{AB}(q)=\hL_{BA}(q)$, only the results for
$\hL_{AA}(q)$ and $\hL_{AB}(q)$ are shown.  

If one assumes that the mobility matrix $\hmatL$ is governed by a single
relaxation time $\tau(q)$ (Eqs.\ (\ref{eq:tauR}) or (\ref{eq:tauE})), the
resulting mobility curves are qualitatively similar to the curves obtained from
the Debye approximation (\ref{eq:Debye}), except that $\hL_{\alpha \beta}(q)$
is enhanced at high $q$ values like the full-chain mobility function.  However,
if one derives $\hmatL$ from a full relaxation time matrix which is calculated
according to Eq.\ (\ref{eq:tauFull}), the mobility functions change
qualitatively.  The intra-block mobility $\hL_{AA}^T(q)$ becomes much larger
than in the other nonlocal schemes, especially at small $q$.  Hence monomer
rearrangements inside blocks are faster than anticipated in the Debye
approximation. Nevertheless, $\hL_{AA}^T(q)$ never reaches the level of the
local coupling scheme, where monomers are taken to move independently
($\hL_{AA}^{\mbox{\tiny Local}}(q)= \hL_{BB}^{\mbox{\tiny Local}}(q) \equiv 0.5
D_c/k_B T$ for symmetric A:B copolymers in homogeneous melts according to Eq.\
(\ref{eq:Local})).

In contrast, the inter-block mobility $\hL_{AB}^T(q)$ is much smaller than
in the other nonlocal schemes already at $q=0$. It then decreases further with
increasing $q$ and even becomes slightly negative, until it rises again and
reaches zero at large $q$. We note that the slightly negative values of
$\hL_{AB}(q)$ do not destabilize the system, since the Eigenvalues of
$\matL(q)$ are still positive.  The inter-block mobility is practically zero
for $q$ values above $q \Rg \approx 1$. The same is obtained with 
a local approximation, where the motion of $A$ and $B$ monomers 
is also uncorrelated.

An important consequence is that the values of $\hL_{AA}(q)$ and $\hL_{AB}(q)$
at $q \to 0$ differ from each other in the relaxation time matrix scheme
$\hmatL^T$ (Eq.\ (\ref{eq:tauFull}), whereas they are equal in the other
nonlocal schemes.  This influences the prediction for the relaxation of
composition fluctuations $m(t) = (\Phi_A(t) - \Phi_B(t))/2$.  From Eq.\
(\ref{eq:DDFT}), one can derive 
\begin{equation}
\partial_t  \: m(\qq,t) = - q^2 \: \frac{1}{2} \: \big(
   \hL_{AA}(\qq) - \hL_{AB}(\qq) \big) \: \hmu(\qq,t),
\label{eq:DDFT_m}
\end{equation}
where $\hmu = (\hmu_A - \hmu_B)$ is conjugate to $m$. If $m(t)$ is small, one
can apply the random phase approximation (RPA)\cite{Fredrickson,Schmid2011} and
approximate $\hmu (\qq,t) \approx  \Gamma_2(\qq) \: m(\qq,t)$,
where the RPA-coefficient $\Gamma_2(\qq)$ can be identified with the inverse
of the collective structure factor of the copolymer melt. Expanding
$\Gamma_2(\qq)$ in powers of $q$ and neglecting compressibility effects,
one obtains to leading order\cite{Schmid2011}
$ \Gamma_2(\qq) \approx {24 \: k_B T}/{q^2 R_g^2}$ for symmetric
diblock copolymers. At small $q$, Eq.\ (\ref{eq:DDFT_m}) thus
takes the limiting form
\begin{equation}
\partial_t  \: m(\qq,t) \approx - \frac{12 k_B T}{R_g^2} \: 
  \big(\hL_{AA}(\qq) - \hL_{AB}(\qq) \big) \: m(\qq,t).
\end{equation}
Since $\big( \hL_{AA}(0) - \hL_{AB}(0) \big) > 0$ in the relaxation time matrix
scheme, composition fluctuations are predicted to decay with a finite
relaxation time in the limit $q \to 0$.  In the other nonlocal schemes, one has
$\big( \hL_{AA}(0) - \hL_{AB}(0) \big) = 0$ at $\qq \to 0$, i.e., the
relaxation time for long-wavelength compositional fluctuations is predicted to
diverge. In simulation studies\cite{Ghasimakbari2019, Wang2019}, the relaxation
time is found to be finite and of order\cite{Ghasimakbari2019} $(2/\pi^2) \:
R_g^2/D_c$ (the Rouse time of the chain), implying $(\hL_{AA}(0)-\hL_{AB}(0))
\approx 0.41 \: D_c/k_BT$. This is consistent with the data in Fig.\
\ref{fig:4} obtained with the relaxation time matrix method. 

\subsection{Comparison of DDFT calculations with simulations}

In order to evaluate the mobility functions discussed in the previous section,
we have compared DDFT calculations with fine-grained simulations for different
situations of dynamical ordering/disordering in block copolymer melts.  In the
following, we report the results for Brownian dynamics simulations. The results
for inertial dynamics simulations are similar.

\subsubsection{Relaxation of an initially lamellar 
symmetric diblock copolymer melt into the homogeneous state}

\begin{figure}[t]
\includegraphics[scale=0.30]{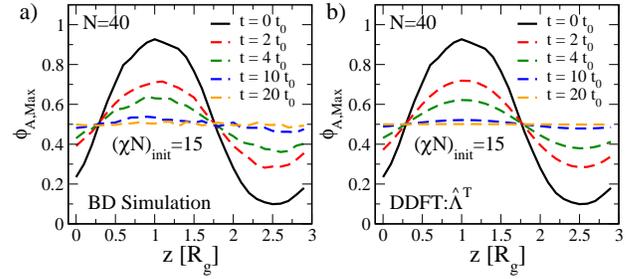}
\caption{Evolution of density profile of A-monomers after a sudden
change from $(\chi N)_{\mbox{\tiny init}} = 15$ to $\chi N = 0$ at $t=0$,
according to (a) Brownian dynamics simulations and (b) DDFT calculations
based on the relaxation time method, Eq.\ (\ref{eq:tauFull}).
}
\label{fig:5}
\end{figure}

In the first example, we study the relaxation of an initially lamellar block
copolymer melt into a homogeneous state.  Diblock copolymer melts were prepared
in a lamellar state by equilibrating them above the order-disorder transition,
i.e., at $(\chi N)_{\mbox{\tiny init}} > (\chi N)_{ODT}$.  Then, starting from
such a configuration, $\chi$ was turned off (to $\chi = 0$) at time $t=0$ and
the evolution of the profiles was monitored.  Fig.\ \ref{fig:5} shows an
example of a series of resulting density profiles for $A$ monomers at different
times, as measured in a Brownian dynamics simulation run (Fig.\ \ref{fig:5}
a)), and the corresponding results from DDFT calculations based on the
relaxation time matrix (Fig.\ \ref{fig:5} b)).  The DDFT calculations are in
excellent agreement with the simulations.

\begin{figure}[t]
\includegraphics[scale=0.3]{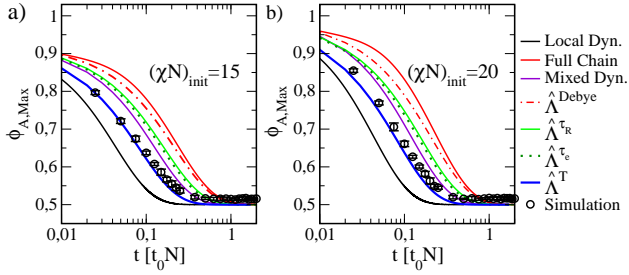}
\caption{Relaxation of the maximum in the density profile of A-monomers 
for configurations that were initially equilibrated in an ordered phase at 
$(\chi N)_{\mbox{\tiny init}}= 15$ (a) 
and $(\chi N)_{\mbox{\tiny init}} = 20$ (b), after a sudden change
to $\chi = 0$ at $t=0$, for different DDFT schemes
as indicated, and compared to Brownian dynamics simulations at $N=40$. 
The initial value $\phi_{A,\mbox{\tiny Max}}(t=0)$ is the same in 
all calculations.
}
\label{fig:6}
\end{figure}

To further quantify the comparison, we plot in Fig.\ \ref{fig:6} the maximum
value of the profile $\Phi_A(z)$ versus time for systems that were initially
prepared at $(\chi N)_{\mbox{\tiny init}}=15$ (Fig.\ \ref{fig:6} a)) and $(\chi
N)_{\mbox{\tiny init}}=20$ (Fig.\ \ref{fig:6} b)). Symbols show the simulation
results, averaged over ten independent runs, and, lines the results from
different DDFT calculations. We find that DDFT calculations based on a
chain coupling assumption (i.e., full chain dynamics, Eq.\
(\ref{eq:FullChain}) or Debye dynamics $\hL^{\mbox{\tiny Debye}}$, Eq.\
(\ref{eq:Debye})), consistently underestimate the speed of the relaxation
process. DDFT schemes with mobility functions $\hL^\tau$ that were
extracted assuming a single relaxation time $\tau(q)$  (i.e., Eqs.\
(\ref{eq:tauR}) and (\ref{eq:tauE})\revision{)} perform better, but the dynamics is
still too slow. The curves calculated with the ''mixed coupling''
scheme\cite{Qi2017}, Eq.\ (\ref{eq:MixDDF}), are close by and also too slow.
DDFT calculations based on a local coupling assumption
overestimate the relaxation speed.  In contrast, the predictions of DDFT
calculations based on the relaxation time matrix, i.e., on $\hL^T$ (Eq.\
(\ref{eq:tauFull})), are in excellent agreement with the simulation data.

\subsubsection{Ordering kinetics in a symmetric diblock copolymer melt}

\begin{figure}[t]
\includegraphics[scale=0.32]{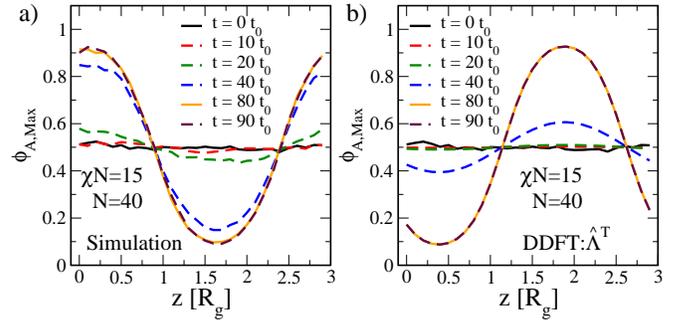}
\caption{Evolution of density profile of A-monomers from an
initially disordered conformation after the monomer interaction
is suddenly raised from $\chi N = 0$ to $\chi N = 15$ at $t=0$,
according to (a) Brownian dynamics simulations and (b) DDFT calculations
with mobility function based on Eq.\ (\ref{eq:tauFull}).}
\label{fig:7}
\end{figure}

In our second example, we study the dynamics of structure formation in the
block copolymer melt after a sudden quench from $\chi N = 0$ to some value
$(\chi N) > (\chi N)_{ODT}$. An example for the time evolution of an A-density
profile obtained from a Brownian dynamics simulation run and compared to DDFT
calculations based on the relaxation time matrix is shown in Fig.\
\ref{fig:7}.  In both cases, the initial density profile is exactly the same,
i.e., small density fluctuations in the simulation profile were also
transferred to the initial configuration in the DDFT calculation. Nevertheless,
the agreement between simulations and DDFT calculations is less impressive than
in the relaxation case, Fig.\ \ref{fig:5}.  First, the location of the density
maxima differs. This can be explained from the fact that the maxima emerge
spontaneously at random positions in both cases. Second, the melt seems to
order faster in the simulations than in the DDFT simulations.  At the time
$t=40 t_0$ after the quench, the amplitude of the oscillation in the
A-density profile has almost saturated in the simulations, whereas it has only
reached about one fourth of the final value in the DDFT calculations.

\begin{figure}[b]
\includegraphics[scale=0.32]{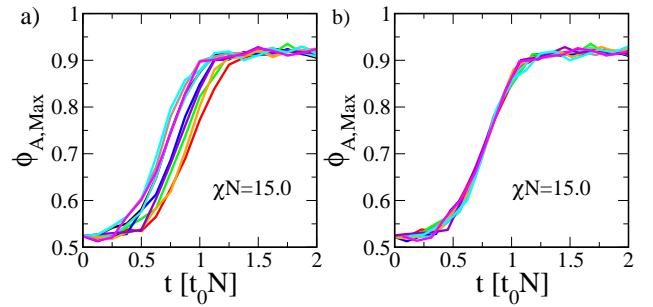}
\caption{(a) Original and (b) aligned curves for the time evolution 
of the maximum in the spatial density of A-monomer after a
sudden quench from $\chi N = 0$ to $\chi N = 15$ at $t=0$ 
from ten different Brownian dynamics simulation runs.}
\label{fig:8}
\end{figure}

On the other hand, looking at the simulations, one notices that the ordering
time also differs between different  simulation runs. Fig.\ \ref{fig:8} shows
results for the maximum value of the A-monomer density profile as a function of
time for ten different independent simulations, which all started from exactly
the same initial configuration at $t=0$.  In every run, the lamellar ordering
sets in at a different time (Fig.\ \ref{fig:8} a)). However, if one aligns the
curves, i.e., adds a time offset such that they coincide at half maximum, their
slopes fall largely on top of each other: The statistical spread of the onset
of the ordering is much larger than the statistical noise after the ordering
has set in. In the following, we therefore not only compare the kinetics 
of ordering on an absolute time scale, but also the shape of the curves after
they have been aligned.

\begin{figure}[t]
\includegraphics[scale=0.31]{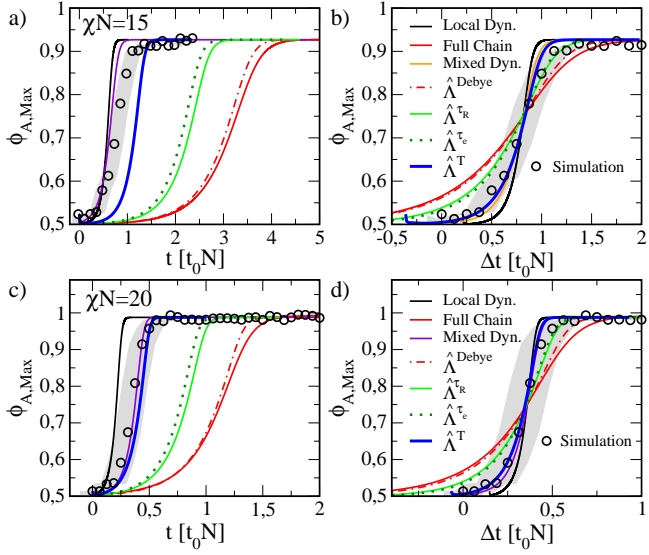}
\caption{(a,c): Time evolution of the maximum density of A-monomers
 after a sudden quench from $\chi N = 0$ to 
 when $\chi N=15.0$ (a) and $\chi N=20.0$ (c), according
 to Brownian dynamics simulations (symbols) and different
 DDFT schemes (lines) as indicated. The initial density profile
 in $z$ direction is the same in all calculations. Grey shades
 indicate spread of simulation curves (see Fig.\ \ref{fig:8}).
 (b,d): Same curves, aligned in time $t$.
}
\label{fig:9}
\end{figure}

Fig.\ \ref{fig:9} shows the corresponding results for quenches to $\chi N = 15$
(Fig.\ \ref{fig:9} a,b), and to $\chi N = 20$ (Fig.\ \ref{fig:9} c,d), compared
to a DDFT predictions from the different schemes discussed above.  As reported
in our earlier work\cite{Qi2017}, and consistent with our observations for the
relaxation kinetics, Fig.\ \ref{fig:6}, DDFT calculations based on local
dynamics (Eq.\ (\ref{eq:Local}, black line) underestimate the ordering time,
and DDFT calculations based on global chain dynamics (full chain dynamics
(\ref{eq:FullChain}) or Debye dynamics (\ref{eq:Debye}, red lines) overestimate
it. Using DDFT mobilities that were extracted from bulk simulations assuming a
single relaxation time, (Eqs.\ (\ref{eq:tauR}) or (\ref{eq:tauE}), green
lines), the predicted ordering is faster than in the case of Debye dynamics,
but still too slow.  

The best results are again obtained with the DDFT scheme $\hL^T$ based on the
relaxation time matrix, Eq. (\ref{eq:tauFull}).  The ordering in the DDFT
calculations sets in later than in the simulations, but once started, the
dynamics of ordering is comparable. The delayed onset may be explained by the
role of thermal fluctuations in initiating the ordering process. The DDFT
calculations are purely deterministic and do not include fluctuations. Since
the initial configurations are chosen identical to the simulated
configurations, they include some noise, and that noise has the correct
amplitude. As we have shown in earlier work\cite{Qi2017}, the ordering would
have been further delayed in all DDFT schemes if the initial noise level had
been chosen lower. Nevertheless, adding noise to the initial configuration of a
deterministic DDFT calculation is apparently not sufficient, if one wishes to
faithfully reproduce the onset of ordering.  To improve on this, one would have
to include thermal noise in the DDFT equations (see Sec.\
\ref{sec:discussion}).  Once initiated, the ordering proceeds in a
deterministic manner and is very well captured by the DDFT calculations based
on $\hL^T$  (Fig.\ \ref{fig:9} b,d, blue line).

The results from "mixed dynamics" calculations (Eq.\ (\ref{eq:MixDDF}), cyan
line) are also in very good agreement with the simulation data. However, it
should be noted that this scheme has been postulated heuristically, without any
microscopic justification, and it has one free parameter (the parameter
$\sigma$ in Eq.\ (\ref{eq:MixFilter})) which has been optimized for this
specific ordering situation in our earlier work\cite{Qi2017}. In contrast, the
mobility functions in the relaxation time scheme were determined from
independent bulk simulations without any adjustable parameter. Also, from
a practical point of view, mixed dynamics calculations have the disadvantage
that they require smaller time steps.

\subsubsection{Asymmetric diblock copolymer melt}

\begin{figure}[t]
\includegraphics[scale=0.31]{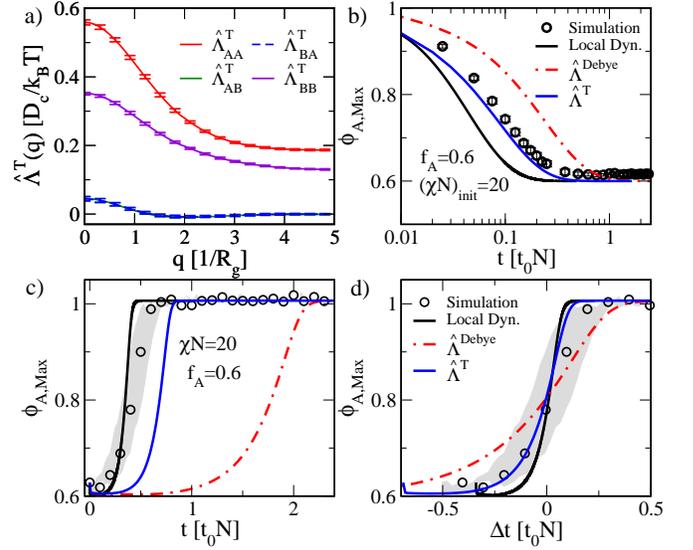}
\caption{(a) Normalized mobility function $\hL^T_{\alpha \beta}$ of
asymmetric A:B diblock copolymers with block ratio 6:4 and length $N=40$
obtained from simulations at $\chi N = 0$ with the relaxation time matrix
approach (Eq.\ (\ref{eq:tauFull})).  (b-d) Corresponding time evolution of the
maximum spatial density of A-monomers (b) after suddenly switching from $\chi N
= 20$ to $\chi N = 0$ (c,d) and from $\chi N = 0$ to $\chi N = 20$ at $t=0$ in
absolute time (c) and aligned in time (d).  Symbols correspond to Brownian
Dynamics simulations with chain length $N=40$, lines to results from DDFT
calculations as indicated. Grey shades indicates spread of simulation curves
from 10 independent configurations with identical starting configuration. }
\label{fig:10}
\end{figure}

So far, we have evaluated our different DDFT schemes by examining systems of
symmetric diblock copolymer melts. To test whether the results depend on the
symmetry of the system, we have repeated the analysis for a different A:B block
fraction.  The results are shown in Fig.\ \ref{fig:10}.  
We consider the same two situations as above: One where an initially lamellar
morphology (set up in the ordered phase at $(\chi N)_{\mbox{\tiny init}}=20$)
relaxes into a homogeneous structure after turning $\chi$ off, and one where an
initially disordered melt develops lamellar order after performing a quench
into the ordered phase at $\chi N = 20$. The results are essentially the same
as in the symmetric case: When using DDFT with ''local dynamics'', the
dynamics is too fast, when using global chain dynamics (Debye dynamics), it is
too slow. When using the relaxation time matrix approach, the onset of ordering
is slightly delayed in the DDFT calculations compared to simulations, but the
actual ordering kinetics (the shape of the curves) is in very good agreement
with the simulation data.

\section{Discussion and Summary}

\label{sec:discussion}

The purpose of the present work was to develop systematic bottom-up
coarse-graining strategies for constructing nonlocal mobility functions
$\hL(\qq)$ in DDFT models for polymeric systems.  The goal was to extract these
mobility functions from trajectories of fine-grained, microscopic simulations.
We have explored two physically motivated approaches.  

The first was based on the Green-Kubo formalism. However, the Green-Kubo
integrals were found to always vanish except at $\qq=0$, due to the fact that
the corresponding stationary current cannot exist at $\qq \ne 0$.  It was
not even possible to identify a well-defined plateau in the running Green-Kubo
integrals.  Espa\~nol et al \cite{Pep2019} have recently discussed 
such ''plateau problems'' and proposed an alternative approach
to evaluating Green-Kubo transport coefficients: They suggested to analyze the
late-time behavior of quantities $- \Big(\frac{\ud}{\ud t} C(t)\Big) \:
C^{-1}(t)$, where $C(t)$ is the time-dependent correlation function of the
quantities of interest. In our case, the relevant correlation function is
the single chain structure factor, $\matG(\qq,t)$. Inserting Eq.\
(\ref{eq:g_DDFT}) yields $\hmatL(\qq) \propto - q^2 (\partial_t \matG(\qq,t))
\matG^{-1}(\qq,t) \: \matG(\qq,0)$.  Since the long-time behavior of
$\matG(\qq,t)$ is dominated by the diffusive behavior of whole chains, one has
$\matG(\qq,t) \propto \exp(- D_c q^2 t)$ at $t \to \infty$ and hence gets
$\matL(\qq) \propto D_c g(\qq,0)$, which corresponds to Debye dynamics. Thus
the resulting DDFT model is a ''chain coupling'' model where chains move as a
whole.  

In practice, however, we are interested in local ordering processes with
characteristic time scales that are typically smaller than the diffusive time.
Therefore, we have explored a second scheme, where a characteristic relaxation
time matrix is first determined independently for each $\qq$-vector from
fine-grained simulations, and this is then used to derive a $\qq$-dependent
mobility matrix.  As one can see from Figs.\ \ref{fig:2} and \ref{fig:4}, the
resulting mobility functions are intermediate between ''chain coupling
dynamics'' (chains move as a whole) and ''local coupling dynamics''
(monomers move independently). We have tested the approach by examining two
kinetic processes in block copolymer melts: The process of disordering from an
initially lamellar phase and the process of ordering after a quench into the
lamellar phase. Comparing the DDFT calculations with the simulation results, we
conclude that our new scheme is capable of describing the ordering/disordering
kinetics at a quantitative level.  
\revision{
Although we applied our model to study the order/disorder kinetics of lamellar 
structures only, the method can be applied to other morphologies as well
(e.g. spheres, cylinders etc.). 
}

We should note that, although the kinetics of ordering and disordering are
well-captured by the DDFT model, the onset of ordering is later than it should
be, compared to simulations. We attribute this to the effect of thermal
fluctuations, which are omitted in our DDFT calculations.  They could be 
included by adding thermal noise to the density currents\cite{Qi2017},
i.e., replace Eq.\ (\ref{eq:DDFT0}) by
\begin{equation}
\partial_t \rho_\alpha =
  \nabla_r \Big\{ \sum_\beta\int \ud \rr' \Lambda_{\alpha\beta}
    \left(\rr,\rr'\right)\nabla_{r'}\mu_\beta
    + \jj_\alpha \big\},
\label{eq:DDFT_noise}
\end{equation}
where the stochastic current $\jj_\zeta(\rr,t)$ is to a Gaussian random vector
field with zero mean ($\langle \jj_\zeta (\rr, t)\rangle = 0$) and
correlations according to the fluctuation-dissipation theorem:
$
\langle j_{{}_I\alpha}(\rr,t) j_{{}_J\beta}(\rr',t') \rangle 
 =  2 k_B T  
   \delta(t-t') \: \Lambda_{\alpha \beta}(\rr,\rr') \: \delta_{{}_{IJ}}
$
($I,J$ are cartesian coordinates).


It is worth recapitulating some of the approximations and assumptions that are
entering our coarse-graining scheme.  

First, we have assumed that the dynamics of inhomogeneous polymer systems can
be described by an effective Markovian model. To account for the multitude of
different relaxation times in polymer systems, we have treated the mobility as
an adjustable $\qq$-dependent function; however, explicit memory effects were
neglected.  Wang et al\cite{Wang2019} have recently devised a dynamic RPA
theory for polymer systems with a frequency dependent Onsager coefficient and
showed that it successfully describes the decay of composition fluctuations in
diblock copolymer melts (similar to Fig.\ \ref{fig:5} here) and the onset of
spinodal decomposition in homopolymer mixtures. Their Ansatz can easily be
generalized to a dynamic SCF theory with a time-dependent memory kernel. It has
the advantage that it includes memory explicitly, and does not require ad hoc
adjustments of ''effective'' mobility functions. On the other hand, effective
Markovian models are computationally more efficient in many cases.

Second, in Eq.\ (\ref{eq:DDFT0}), the mobility matrix describing the time
evolution of density fluctuations should really be derived from the {\em
collective} density correlations. Here, we have replaced them by a sum over
{\em intrachain} density correlations, in the spirit of a mean-field theory.
Recently, Ghasimakbari and Morse\cite{Ghasimakbari2019} have used the
collective structure factor to analyze the effective $\qq$-dependent
diffusive relaxation of compositional fluctuations in symmetric diblock
copolymer melts.  They fitted the decay of the dynamic collective structure
factor by a single exponential. Their results in the regime $(\chi N) < 10.5$
are comparable to ours in Fig.\ \ref{fig:3}. 

Third, when deriving our final expression for $\matL(\qq)$ in Eq.\
(\ref{eq:fs}), we have linearized the free energy density functional and thus
assumed that density variations are small. We determine the mobility function
$\hL(\qq)$ from simulations of a homogeneous bulk melt at $\chi N = 0$, but
then use them in DDFT calculations for inhomogeneous, ordered systems.
This is partly motivated by the finding that $\hL(\qq)$ hardly depends on
$\chi$ in the disordered regime of a block copolymer melt. Nevertheless, at
high $\chi$ and/or in strongly inhomogeneous systems, corrections must probably
be applied.

We have formulated our approach for diblock copolymer melts, but it can
easily be generalized to mixtures. Starting from Eq.\ (\ref{eq:DDFT0}), 
one can simply replace the mobility function 
$\Lambda_{\alpha \beta} \approx \Lambda_{\alpha \beta}^{(s)} \: \rho_0/N$,
by a sum over chain mobilities, i.e.
\begin{equation}
\label{eq:DDFT_mixture_1}
\Lambda_{\alpha \beta}(\rr,\rr')
  = \sum_\gamma \frac{1}{N_\gamma} \bar{\rho}^{(\gamma)}(\rr,\rr') \:
    \Lambda_{\alpha\beta}^{(s,\gamma)}(\rr,\rr')
\end{equation}
where the sum $\gamma$ runs over chain types, $N_\gamma$ is the length
of chains of type $\gamma$, $\bar{\rho}^{(\gamma)}(\rr,\rr')$ the
locally averaged density of monomers from chains of type $\gamma$
(hence $\bar{\rho}^{\gamma}/N_\gamma$ is a chain density), 
and $\Lambda_{\alpha \beta}^{(s,\gamma)}$ the corresponding single chain 
mobility function. Note that the prescription for determining the local
average $\bar{\rho}^{(\gamma)}(\rr,\rr')$ must be symmetric 
with respect to $\rr$ and $\rr'$ (e.g.,
$\bar{\rho}^{(\gamma)}(\rr,\rr') = \rho^{(\gamma)}(\frac{\rr+\rr'}{2})$.



In mixtures, the diffusion of chains of different type adds another slow time
scale to the dynamics of the system.  In our previous work\cite{Qi2017}, we
have compared the dynamics of interdiffusion at A/B homopolymer interfaces
from different DDFT calculations with simulations. We found that
the results obtained with local and nonlocal DDFT coupling schemes
were very similar, and all in very good agreement with the simulations.
We conclude that studies of homopolymer interdiffusion do not seem to be
a very sensitive test of the quality of a DDFT model, and therefore expect that
the new schemes proposed here will also perform well.

Our bottom-up approach for constructing mobility matrices has been tested for
Rouse chains, but it is not restricted to that.  It only requires as
input the single chain dynamic structure factors from simulations of the target
microscopic systems. In future work, we plan to study polymer mixtures and
melts in other dynamical regimes, e.g., entangled melts, or systems where
hydrodynamics are important. 

The DDFT theory relies on the assumption that the polymer system under
consideration is only weakly disturbed from equilibrium. It assumes that the
polymer conformations are close to local equilibrium at all times
\revision{and that the dynamic process under consideration is still
suitably described in terms of a free energy landscape picture.} Therefore,
it cannot be applied in situations far from equilibrium where the distribution
of polymer conformations is distorted, such as, e.g., polymers under shear at
high Weissenberg numbers which are stretched out.  Studying such systems with
DDFT models requires novel approaches where not only the mobility functions,
but also the density functionals themselves have to be reconsidered
\cite{Mueller2015, Chandran2019}. However, DDFT theories that were constructed
as proposed in the present paper can be used to study ordering processes and
spontaneous self-assembly in inhomogeneous polymer mixtures, and thus to
evaluate the role of processing and pathways for the final structures.

\subsection*{Acknowledgements}
We thank Marcus M\"uller for a critical reading of the manuscript and many
useful comments.  This research was supported by the German Science Foundation
(DFG) via SFB TRR 146 (Grant number 233630050, project C1). S.Q. acknowledges
research support from the National Natural Science Foundation of China under
the Grant NSFC-21873010.  The simulations were carried out on the high
performance computing center MOGON at JGU Mainz.

\begin{appendix}

\revision{
\section{Evaluation of the Green-Kubo integral}
}

\label{app:green_kubo}

\revision{In this appendix, we discuss the results from the evaluation
of the integral (\ref{eq:GK}).} In the spirit of mean-field theory,
we will assume that the mobility can be derived from a single chain mobility,
$\Lambda = \frac{\rho_0}{N} \Lambda^{(s)}$, which is derived from the
current-current correlations of a single chain, i.e., the quantity
\begin{equation}
\label{eq:II}
\II_{\alpha \beta}(\qq,t) =
\sum_{k,j=1}^N e^{i \qq \cdot (\RR_k(t) - \RR_j(0)) } \:
  \dot{\RR}_k(t) \dot{\RR}_j(0) \: \gamma_k^{(\alpha)} \gamma_j^{(\beta)}.
\end{equation}
If interchain correlations can be neglected, one has $\langle
\jj_\alpha(\qq,t) \jj_\beta(-\qq,t) \rangle = n_c \II_{\alpha \beta} (\qq,t)$,
where \mbox{$n_c = V \frac{\rho_0}{N}$} is the number of polymers in the
system, and hence
\begin{equation}
\LGK_{\alpha \beta}(\qq) = \frac{1}{k_B T} 
\int_0^\infty \!\!\!\! \ud t \: \II_{\alpha \beta}(\qq,t) : \hq \hq.
\end{equation}
The full chain mobility (all monomers) is given by the sum
$\Lambda^{(s)} (\qq) = 
\sum_{\alpha \beta} \Lambda^{(s)}_{\alpha \beta} (\qq)$. 

We first discuss the full chain mobility at $\qq = 0$.  Eq.\ (\ref{eq:II}) then
reduces to $\II(0,t) = \sum_{\alpha \beta} \II_{\alpha \beta}(0,t) = \sum_{kj}
\langle \dot {\RR}_k(t) \dot{\RR}_j(0) \rangle$. After evaluating the average
of $\hq \hq$ with respect to all possible directions $\hq$, we recover the
well-known relation between the chain mobility and the velocity autocorrelation
function of the center of mass of the chain
($\VV(t) = \frac{1}{N} \sum_k \dot{\RR}_k(t)$): 
\begin{equation}
\LGK(0) = \frac{N^2}{3 k_B T} \int_0^\infty \ud t
\left< \VV(t) \VV(0) \right> = \frac{D_c N^2}{k_B T}.
\end{equation}
Here $D_c$ is the diffusion constant of the whole chain, and the factor
$N^2$ accounts for the fact that $\Lambda^{(s)}$ describes the response
of monomer current (scaling with the number $N$ of monomers) to a thermodynamic
force acting on monomers (i.e., the total force again scales with $N$).

For $\qq \neq 0$ and $t > 0$, $\II_{\alpha \beta}(\qq,t)$ can be derived 
from the single chain dynamic structure factor, defined as \cite{Doi}
\begin{equation}
g_{\alpha \beta}(\qq,t) = \frac{1}{N}
\left< \sum_{k,j=1}^N e^{i \qq \cdot(\RR_k(t) - \RR_j(0))}
\gamma_k^{(\alpha)} \gamma_j^{(\beta)} \right>
\end{equation}
by taking the second derivative with respect to $t$:
\begin{equation}
\II_{\alpha \beta}(\qq,t):\qq \qq
= - N \frac{d^2}{dt^2} g_{\alpha \beta}(\qq,t). 
\end{equation}
Putting everything together, we finally obtain the following Green-Kubo
relation between the mobility function and the single chain dynamic structure
factor,
\begin{eqnarray}
\label{eq:GKGP}
\LGK_{\alpha\beta}\left(q\right) 
&= &\frac{N}{k_B T }
\left(\frac{1}{q^2} \lim_{t\to0}\left[\frac{d}{dt}
   g_{\alpha\beta}\left(q,t\right)\right] \right.
\\ && \quad
\nonumber \left.
  + \: \lim_{\epsilon \to 0}
 \int_0^\epsilon \ud t \:  \II_{\alpha \beta}(\qq): \hq \hq \right).
\end{eqnarray} 
This quantity can be measured in microscopic simulations. 
The second term in Eq.\ (\ref{eq:GKGP}) has to be added explicitly 
if the microscopic model evolves according to overdamped Brownian dynamics, 
to account for the contribution of the delta-correlated stochastic 
white noise at $t=0$ to Eq.\ (\ref{eq:GK}).

\begin{figure}[b]
\includegraphics[scale=0.32]{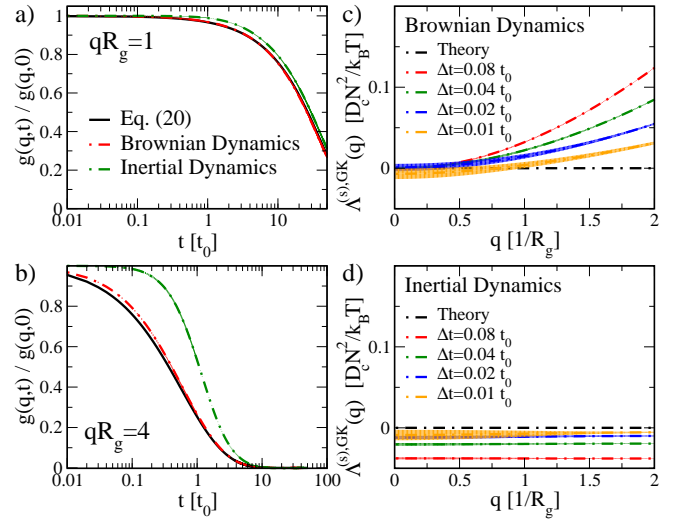}
\caption{
Left: Normalized single chain dynamic structure factor of homopolymers 
with length $N=40$ in a homopolymer melt, as obtained from Brownian
dynamics (red) and inertial dynamics (green) simulations at $q\Rg=1.0 $ (a) 
and $q\Rg=4.0$ (b).  Black line shows the analytical prediction of
Eq. (\ref{eq:Doi}).  
Right: Normalized mobility function obtained via the Green-Kubo
relation (\ref{eq:GKGP}) from Brownian dynamics (c) and 
inertial dynamics (d) simulations. The derivatives of $g(q,t)$ were
taken numerically using a forward difference scheme with different 
values of $\Delta t$ as indicated. The units $t_0$ and $\Rg$ are 
simulation units (see text).  }
\label{fig:1}
\end{figure}

Fig.\ \ref{fig:1} shows simulation results for single chains in a homogeneous
melt from Brownian dynamics and inertial dynamics simulations (see Sec.\
\ref{sec:simulations} for a detailed description of the simulation models).
Fig.\ \ref{fig:1} a,b) show results for $g(\qq,t)$ for $q R_g = 1$ (a) and $q
R_g = 4$ (b) and compares them with an analytic result for ideal free 
Rouse chains\cite{Doi}, which is exact in the limit $N \to \infty$:
\begin{eqnarray}
\nonumber
g\left(q,t\right) & = &
\frac{1}{N}\sum_{ij}\exp\left[ -q^2 D_c t - \frac{|i-j| (q\Rg)^2}{N}  \right.
\\ \nonumber & & 
- \frac{4 (q \Rg)^2}{\pi^2}\sum^N_{p=1}
\frac{1}{p^2}\cos\left(\frac{p\pi i}{N}\right)
\cos\left(\frac{p\pi j}{N}\right)
\\ & & \left.  \qquad \qquad
\left\{ 1 - \exp\left(-\frac{D_c t p^2\pi^2}{2\Rg^2} \right)\right\}\right].
\label{eq:Doi}
\end{eqnarray}
Here, the index $p$ represents the $p$th Rouse mode, and the indices $i,j$
represent the $i$th and $j$th beads on the polymer chain. The agreement with
the Brownian dynamics simulaton data is very good.
Fig.\ \ref{fig:1} c,d) shows the corresponding Green-Kubo mobility functions.
Somewhat disappointingly, they are found to be zero within the statistical and
systematic error.  Deviations from zero can be traced back to discretization
artefacts when taking the derivative $\frac{d}{dt} g(q,t)$ numerically.

In the case of overdamped Rouse homopolymers, we can evaluate
(\ref{eq:GKGP}) exactly, using the relation \cite{Doi}
\begin{eqnarray}
\frac{d}{dt} \ln g(q,t) &=& - \frac{1}{g(q,0)} \frac{k_B T}{N}
\\ \nonumber && \quad  
\sum_{kj} \left< \mathbf{H}_{kj} 
  \exp( i \qq \cdot (\RR_k - \RR_j)) \right> : \qq \qq
\end{eqnarray}
with the Rouse mobility matrix $\mathbf{H}_{kj} = D_0 \mathbf {1} \delta_{kj}$.
The first term in (\ref{eq:GKGP}) yields $\frac{1}{q^2} \frac{d}{dt}
g(q,t)\big|_{t\to0} = - D_0 N k_B T$.  The noise term contributes with $2 k_B T
D_0 N \int_0^\epsilon \ud t \delta(t) = D_0 N k_B T$.  Since these two terms
cancel, the resulting Green-Kubo mobility is zero, as suggested by the
simulations.  

\end{appendix}

\bibliography{LitKOnsg}

\end{document}